\documentclass{article}

\usepackage[preprint,nonatbib]{neurips_2020}

\usepackage[utf8]{inputenc} 
\usepackage[T1]{fontenc}    
\usepackage{hyperref}       
\usepackage{url}            
\usepackage{booktabs}       
\usepackage{amsfonts}       
\usepackage{nicefrac}       
\usepackage{microtype}      

\usepackage{amsmath, amssymb, amsfonts, amsthm}
\usepackage{graphicx}
\usepackage{url}
\usepackage{bm}
\usepackage{subfigure}
\usepackage{graphicx}
\usepackage{mathabx}
\usepackage{multirow}
\usepackage{setspace}
\usepackage[separate-uncertainty = true,multi-part-units=single]{siunitx}
\usepackage{biocon}
\usepackage{booktabs}
\usepackage{outlines}
\usepackage{textcomp}
\usepackage{tabularx}
\usepackage{makecell}
\usepackage{rotating}
\usepackage{lscape} %
\usepackage{longtable} %
\usepackage[roman]{parnotes}
\usepackage[version=4]{mhchem}
\usepackage{contour} 
    \contourlength{0.8pt}
\usepackage{verbatim}
\usepackage{booktabs}
\usepackage{multirow}
\usepackage[table,xcdraw,dvipsnames]{xcolor}
\usepackage{caption}
\usepackage{float} 
\usepackage{array} 
\usepackage{longtable}
\usepackage{csvsimple}
\usepackage{pdfpages}
\usepackage{longtable}
\usepackage{authblk}

\title{Use HiResCAM instead of Grad-CAM for faithful explanations of convolutional neural networks}

\author[a]{Rachel Lea Draelos Ph.D.}
\author[b]{Lawrence Carin Ph.D.}
\affil[a]{Corresponding author. rlb61@duke.edu. Duke University Computer Science}
\affil[b]{lawrence.carin@kaust.edu.sa. Duke University Electrical and Computer Engineering}

\begin{document}

\maketitle 

\begin{abstract}

Explanation methods facilitate the development of models that learn meaningful concepts and avoid exploiting spurious correlations. We illustrate a previously unrecognized limitation of the popular neural network explanation method Grad-CAM: as a side effect of the gradient averaging step, Grad-CAM sometimes highlights locations the model did not actually use. To solve this problem, we propose HiResCAM, a novel class-specific explanation method that is guaranteed to highlight only the locations the model used to make each prediction. We prove that HiResCAM is a generalization of CAM and explore the relationships between HiResCAM and other gradient-based explanation methods. Experiments on PASCAL VOC 2012, including crowd-sourced evaluations, illustrate that while HiResCAM's explanations faithfully reflect the model, Grad-CAM often expands the attention to create bigger and smoother visualizations. Overall, this work advances convolutional neural network explanation approaches and may aid in the development of trustworthy models for sensitive applications.

\end{abstract}

\section{Introduction}

\begin{figure}
    \centering
    \includegraphics[scale=0.08]{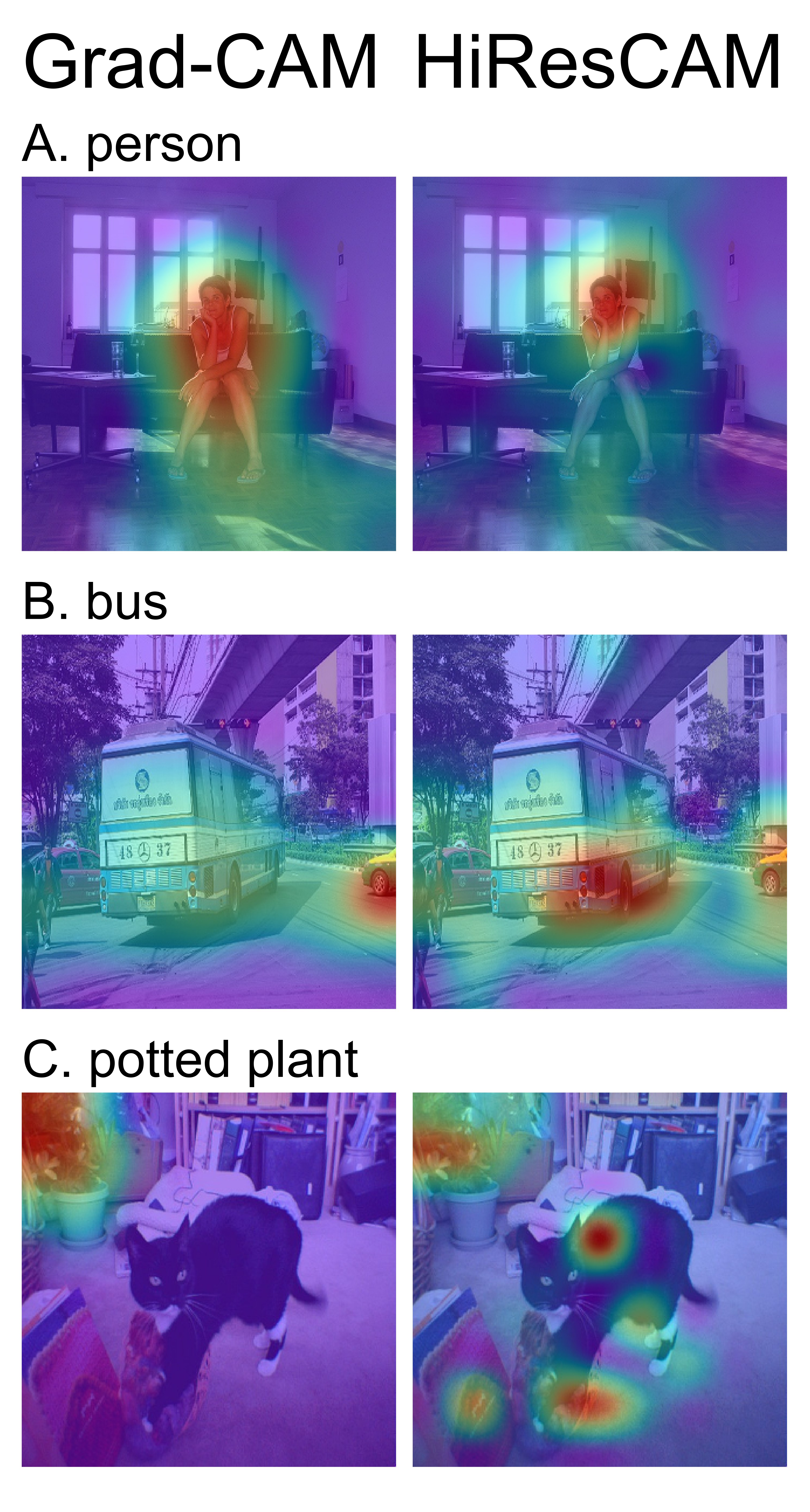}
    \caption{Grad-CAM and HiResCAM produce different explanations for the same model, image, and class. HiResCAM provably reflects the locations the model used for computation, while Grad-CAM does not. (A) HiResCAM explanations are often more focal than Grad-CAM explanations. (B) Sometimes HiResCAM highlights the correct object while Grad-CAM does not; (C) other times, HiResCAM highlights more parts of the image than Grad-CAM. In this example, the Grad-CAM explanation gives the impression that the model predicted ``potted plant'' from the plant alone, but the HiResCAM explanation reveals that the model also attended to other parts of the image. Best viewed in color.}
    \label{fig:specificpascal}
\end{figure}

Machine learning models sometimes rely on spurious correlations, predicting horses from copyright watermarks \cite{lapuschkin2019unmasking} or pneumonia from metal tokens \cite{zech2018variable}. Even more alarming, models absorb prejudices present in training data and exhibit racial and gender bias \cite{najibi2020racial, buolamwini2018gender}. Explanation methods are one tool to identify and combat these dangerous properties. By improving understanding of how a model makes predictions, explanation methods can facilitate selection of more reasonable, fair models. For convolutional neural networks (CNNs) that process images, visual explanation methods are often used. These methods highlight the parts of an input image that contributed most to a particular prediction.

Gradient-based visual explanation methods for CNNs are popular due to their computational efficiency. Input-level gradient-based explanation methods \cite{simonyan2014deep, zeiler2014visualizing, springenberg2014striving, shrikumar2016not, bach2015LRPoriginal} are quick, but produce explanations that are not class-specific in practice due to a ``white noise'' appearance. Output-level gradient-based methods such as Class Activation Mapping (CAM) \cite{zhou2016CAM} and Grad-CAM \cite{selvaraju2017grad} are class-specific, and thus have been deployed in numerous settings including sensitive applications like medical imaging \cite{lee2019explainable, panwar2020deep, baltruschat2019comparison, shen2018dynamic, pasa2019efficient, lu2019deep}. Grad-CAM in particular is often used because its explanations can be calculated for any CNN architecture, while CAM is restricted to a subset of CNN architectures.

We identify a previously unreported limitation of Grad-CAM: due to its gradient averaging step, Grad-CAM is not guaranteed to reflect locations the model used for prediction, and therefore can produce misleading explanations. To solve this problem, we develop HiResCAM, an explanation method in the CAM family that is faithful to the model. Grad-CAM and HiResCAM explanations often differ noticeably as exemplified in Figure \ref{fig:specificpascal}. In this work, we offer the following contributions:

\begin{itemize}
    \item We develop HiResCAM, a novel explanation method, and prove that HiResCAM explanations are guaranteed to reflect the locations the model used for any CNN ending in one fully connected layer, while the same guarantee does not hold for Grad-CAM;
    \item We prove that HiResCAM is a generalization of CAM and conceptually connect HiResCAM to other gradient-based explanation methods;
    \item In experiments on natural images, we quantify how Grad-CAM explanations deviate from the model's calculations, and show through examples and crowd-sourced assessment that Grad-CAM's explanations are often bigger and rounder than the faithful HiResCAM explanations.
\end{itemize}

\section{Related Work}

Our proposed attention mechanism, HiResCAM, is part of the family of gradient-based neural network explanation methods \cite{ancona2019XAIBook}.

\subsection{Input-level approaches}

Saliency mapping \cite{simonyan2014deep}, DeconvNets \cite{zeiler2014visualizing}, and Guided Backpropagation \cite{springenberg2014striving} are gradient-based explanation methods that compute the gradient of the class score with respect to the input image to visualize important image regions. These methods are identical except for handling of ReLU nonlinearities \cite{nie2018theoretical}. Gradient $*$ Input \cite{shrikumar2016not} is a related method in which the gradient of the class score with respect to the input is multiplied element-wise against the input itself. Layer-wise relevance propagation (LRP) \cite{bach2015LRPoriginal} proceeds layer-by-layer, starting with the output, to redistribute the final score across the pixels of the input. While not originally formulated as a gradient-based explanation method, $\epsilon$-LRP is in fact equivalent to Gradient $*$ Input where the gradient is calculated in a modified manner using the ratio between the output and input at each nonlinearity \cite{ancona2019XAIBook}. A limitation of the aforementioned approaches is the ``white noise'' appearance of the final explanation caused by shattered gradients \cite{balduzzi2017shattered, nie2018theoretical}, which prevents the explanation from being class-specific in practice.

\subsection{Output-level approaches} 

Class Activation Mapping (CAM) \cite{zhou2016CAM} is an explanation method for a particular class of neural networks that consist of convolutions followed by global average pooling of feature maps and one final fully connected layer. CAM explanations are produced by multiplying the class-specific weights of the final layer by the corresponding feature maps prior to pooling. CAM may be considered a gradient-based method, as the final class-specific weights are the gradient of the score with respect to the feature maps. Grad-CAM \cite{selvaraju2017grad} is a generalization of CAM, in which gradients are averaged over the spatial dimensions to produce importance weights. The Grad-CAM explanation is a sum of feature maps weighted by the importance weights. Grad-CAM was proposed with the intention of extending CAM explanations to a broader class of CNN architectures.

Unlike the input-level approaches, which rely on propagation through all layers back to the level of the original image, CAM and Grad-CAM produce explanations at a layer of the network closer to the output. The low-dimensional explanation is then upsampled for superimposition over the input image, an acceptable step because typical CNNs preserve spatial relationships. Guided Grad-CAM \cite{selvaraju2017grad} is a Grad-CAM variant obtained via an element-wise product of the Guided Backpropagation \cite{springenberg2014striving} and Grad-CAM explanations.

Recent work has called into question some gradient-based explanation methods. Nie \textit{et al.} \cite{nie2018theoretical} demonstrate that Guided Backpropagation and DeconvNets perform partial image recovery due to their handling of ReLU nonlinearities and max pooling. Adebayo \textit{et al.} \cite{adebayo2018sanity} reveal that Guided Backpropagation, Guided Grad-CAM, and Gradient $*$ Input produce convincing explanations even when model parameters have been randomized or when a model has been trained on randomly labeled data; however, saliency mapping and Grad-CAM pass their sanity checks. 

Because they are are class-specific and produce less noisy explanations than input-level methods, CAM and Grad-CAM form the foundation of numerous weakly supervised localization methods \cite{kolesnikov2016SEC, wang2018MCOF, huang2018DSRG, wei2018MDC, jiang2019integral, wang2020SEAM, fan2020ICD,ahn2018affinitynet,fan2020cian, chang2020mixupcam, chang2020SCCAM, li2018GAIN, lee2019ficklenet}. CAM and Grad-CAM have also been widely used to explain model behavior, including in sensitive medical applications where protecting patients from biased, unreasonable models is particularly important \cite{lee2019explainable, panwar2020deep, baltruschat2019comparison, shen2018dynamic, pasa2019efficient, lu2019deep}. Unfortunately, as we will demonstrate, Grad-CAM is not a reliable explanation method and sometimes highlights locations the model did not use. We thus propose a new explanation method, HiResCAM, and prove that HiResCAM explanations faithfully reflect the model's computations.

\section{Methods}

\subsection{Problem setup}

Consider a CNN that takes in an input image $\mathbf{X}$. The goal of class-specific visual model explanation is to produce an attention map for a class $m$ with the same shape as input image $\mathbf{X}$ and values in $[0,1]$ such that higher values indicate regions of the input image that increase the model's score for class $m$. 

An output-level CNN explanation approach accomplishes this goal by applying the trained CNN with fixed parameters to an input image $\mathbf{X}$, in order to obtain an explanation at the level of some convolutional feature maps. The explanation is then upsampled and superimposed over $\mathbf{X}$ for visualization.

\subsection{Motivation}  \label{section:hirescam}

We propose a new output-level, gradient-based CNN explanation approach, High-Resolution Class Activation Mapping (HiResCAM), which produces for every class $m=1,...,M$ a class-specific attention map $\mathcal{\tilde{A}}_m^{\rm{HiResCAM}} \in \mathbb{R}^{D_1 \times D_2}$, \textit{i.e.} attention over the spatial dimensions $D_1$ and $D_2$ excluding the feature dimension $F$. 

HiResCAM is inspired by the popular attention mechanism Grad-CAM \cite{selvaraju2017grad}, and is designed to address a limitation of the Grad-CAM averaging step. In Grad-CAM, feature map importance weights $\boldsymbol{\alpha}^f$ are calculated by averaging gradients over the spatial dimensions of the low-dimensional CT representation. Such averaging is likely motivated by the global average pooling step built in to the architecture of Class Activation Mapping (CAM) \cite{zhou2016CAM}. However, the averaging limits the extent to which the final visualization depicts the locations within the image that the model is using to make predictions.

\begin{figure}
\begin{center}
    \includegraphics[scale=0.5]{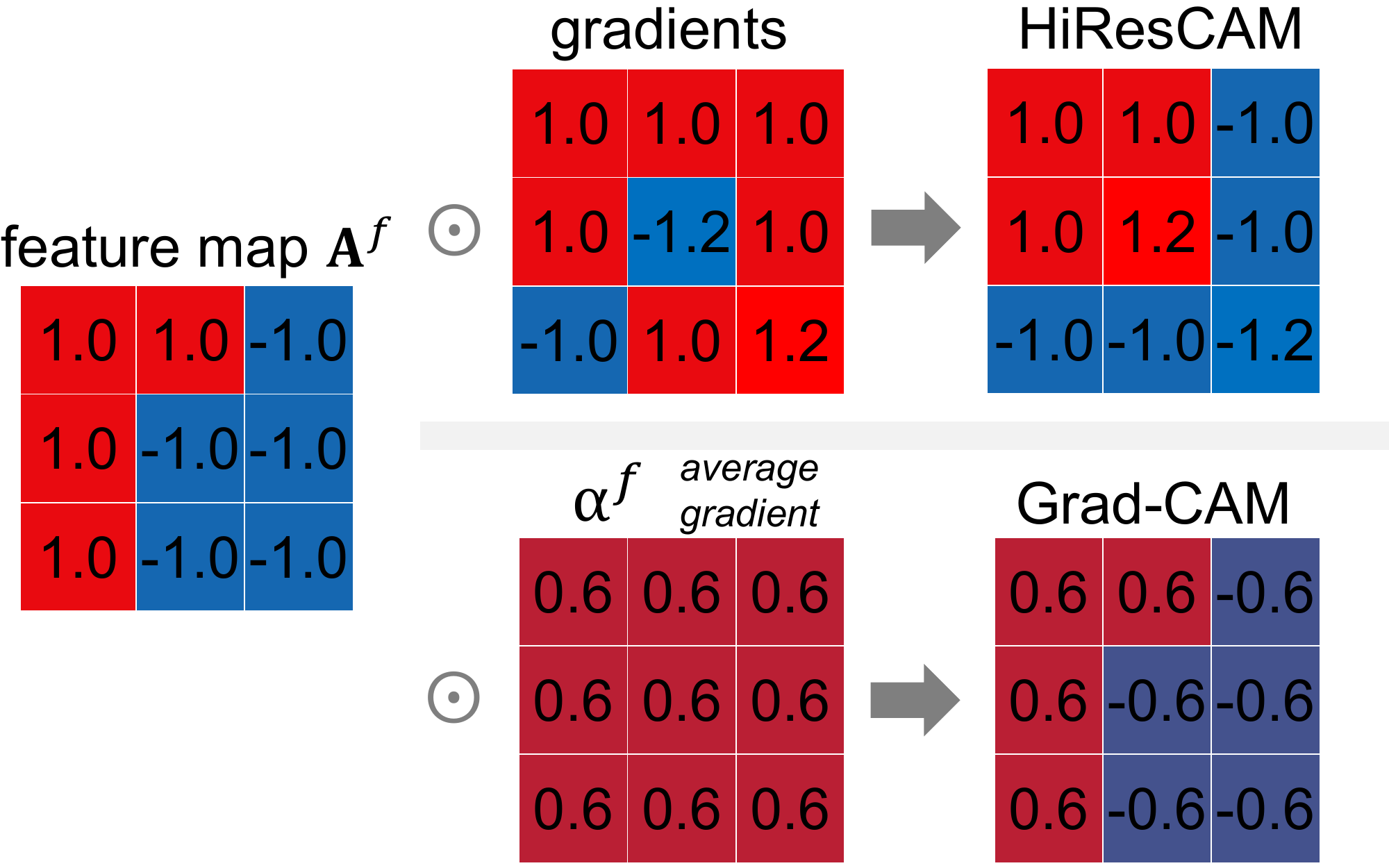}
    \caption{2D example of how HiResCAM addresses the limitation of the gradient averaging step in Grad-CAM. The Grad-CAM explanation (equation \ref{eqn:vanillagradcam}) matches the relative magnitudes and positive-negative pattern of the original feature map (the ``inverted red L shape'' here), even though the gradients suggest that some elements should be re-scaled and/or change sign. HiResCAM (equation \ref{eqn:hirescam}) does not average over the gradients and instead element-wise multiplies the feature map with the gradients directly, thereby producing attention that reflects the model's computations and emphasizes the most important locations for a particular prediction. Best viewed in color.}
    \label{fig:hirescam}
\end{center}
\end{figure}

Figure \ref{fig:hirescam} illustrates the fundamental problem: each $\alpha^f \mathbf{A}^f$ subcomponent of the final Grad-CAM explanation must always match the relative magnitudes of the feature map $\mathbf{A}^f$, and either (a) exactly match the positive-negative pattern of the feature map (when $\alpha^f$ is positive), or (b) invert the positive-negative pattern (when $\alpha^f$ is negative). Rescaling and sign changes of individual elements of the feature map are ``blurred out.'' In HiResCAM, rescaling and sign changes are preserved, producing more high-resolution attention that reflects the model's computations.

\subsection{Grad-CAM formulation}

Before specifying the HiResCAM formulation, we review how Grad-CAM explanations are calculated. Define $s_m$ as a CNN's raw score for class $m$ before a sigmoid or softmax function is applied to produce predicted probabilities. To obtain a Grad-CAM explanation for class $m$, we first compute the gradient of $s_m$ with respect to a collection of feature maps $\mathbf{A}= \{ \mathbf{A}^f \}_{f=1}^F$ produced by a convolutional layer. For 2D data, this gradient $\frac{\partial s_m}{\partial \mathbf{A}}$ is 3-dimensional, $[F, D_1, D_2]$, matching the shape of the collection of feature maps. Note that the collection of feature maps selected could be produced by the last convolutional layer, or an earlier convolutional layer, as the authors of Grad-CAM state that Grad-CAM explanations can be calculated at any convolutional layer \cite{selvaraju2017grad}.

After computing $\frac{\partial s_m}{\partial \mathbf{A}}$, we calculate a vector of importance weights \cite{selvaraju2017grad} $\boldsymbol{\alpha}_m \in \mathbb{R}^F$, where each element $\alpha_m^f$ will be used to re-weight the corresponding feature map $\mathbf{A}^f$. The importance weights are obtained by global average pooling the gradient over the spatial dimensions:
\begin{align}
    \alpha_m^f = \frac{1}{D_1 D_2} \sum_{d_1=1}^{D_1} \sum_{d_2=1}^{D_2} \frac{\partial s_m}{\partial \mathbf{A}^f_{d_1 d_2}}. \label{eqn:importanceweights}
\end{align}
The importance weights indicate which features are most relevant to this particular class throughout the image overall. The final Grad-CAM explanation is produced as an importance-weighted combination of the feature maps:
\begin{align} \label{eqn:vanillagradcam}
    \mathcal{\tilde{A}}_m^{\rm{GradCAM}} = \sum_{f=1}^F \alpha_m^f \mathbf{A}^f.
\end{align}
Following standard practice for use of Grad-CAM \cite{selvaraju2017grad}, the attention map is then post-processed for better visualization by applying a ReLU and normalizing the attention values to the range $[0,1]$. This step ensures that the regions positively associated with a class will be easily visible.

\subsection{HiResCAM formulation}

HiResCAM addresses the limitation of Grad-CAM illustrated in Figure \ref{fig:hirescam}. The first step of HiResCAM is the same as the first step of Grad-CAM: compute $\frac{\partial s_m}{\partial \mathbf{A}}$, the gradient of $s_m$ with respect to the feature maps $\mathbf{A}$.

In the second step of HiResCAM, the attention map is produced by element-wise multiplying the gradient and the feature maps before summing over the feature dimension:
\begin{align} \label{eqn:hirescam}
    \mathcal{\tilde{A}}_m^{\rm{HiResCAM}} = \sum_{f=1}^{F} \frac{\partial s_m}{\partial \mathbf{A}^f} \odot \mathbf{A}^f.
\end{align}
The purpose of HiResCAM is to reflect the model's computations; therefore, when the gradients indicate that some elements of the feature map should be scaled or have their sign inverted, HiResCAM performs these operations. In contrast, Grad-CAM blurs the effect of the gradients across each feature map.

HiResCAM can be applied to 2D, 3D, or $n$-D CNNs; the formulation works for feature maps with any number of spatial dimensions.

\subsection{CNNs ending in one fully connected layer} \label{sec:cnn-end-one-fc}

\subsubsection{HiResCAM highlights locations that increase the class score} \label{section:hirescamproof}

HiResCAM can be calculated at any convolutional layer of a CNN. However, for an explanation with guaranteed properties, HiResCAM must be applied at the last convolutional layer of any CNN that ends in one fully connected layer, which includes many modern CNN architectures. In this section we prove that for such CNNs, HiResCAM has an intuitive interpretation: the resulting explanation is guaranteed to highlight all locations within the image that increase the class score. In contrast, Grad-CAM's explanations do not reflect calculation of the class score.

\begin{figure}
    \centering
    \includegraphics[scale=0.44]{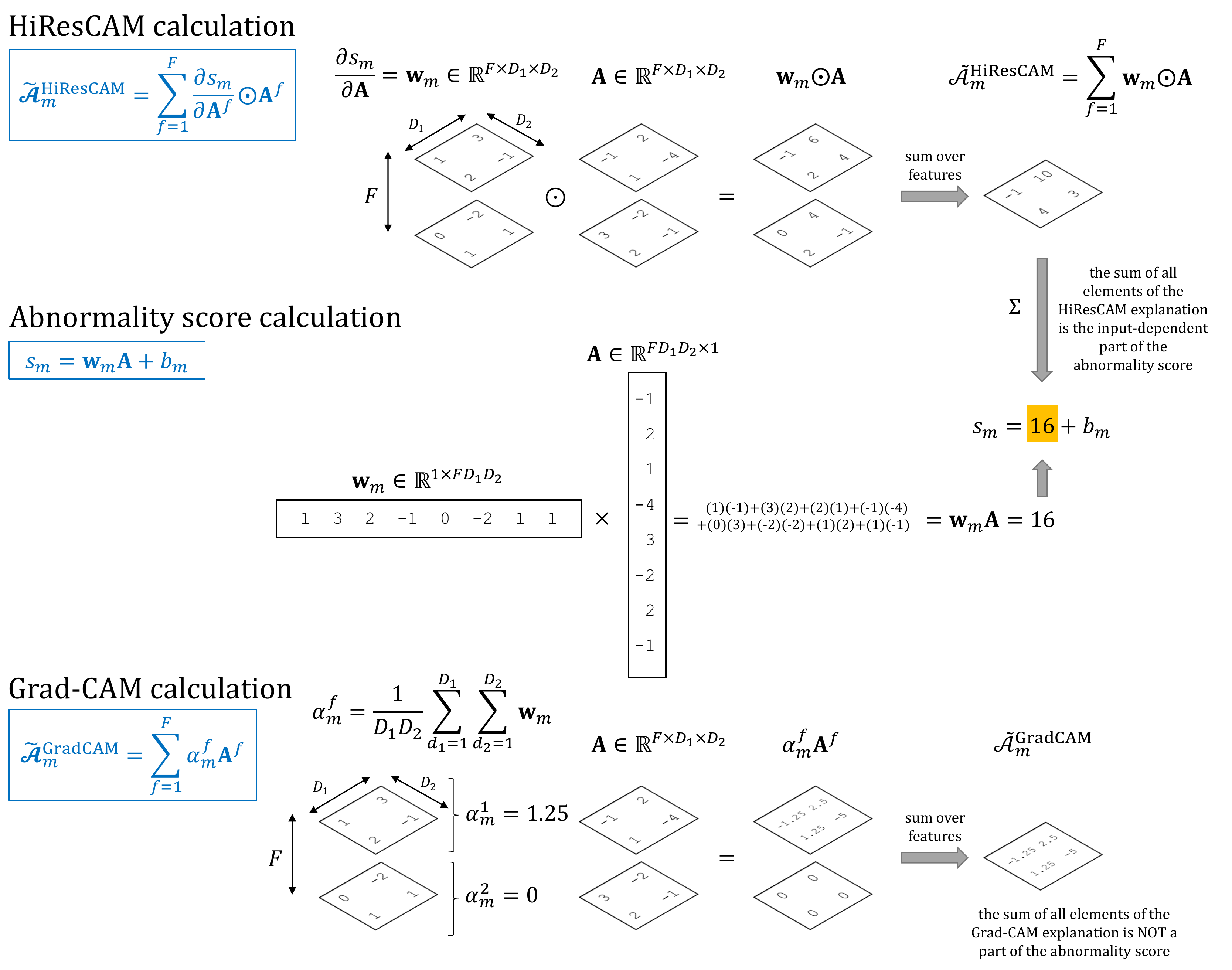}
    \caption{Specific example demonstrating that for CNNs ending in a single fully connected layer, HiResCAM explanations directly reflect the calculation of the class score while Grad-CAM explanations do not. Integer input values were chosen for simplicity; actual weights and activations are not integers.}
    \label{fig:hirescamgradcamconcrete}
\end{figure}

\begin{proof}

Consider a 2D CNN constructed from the following layers:

\begin{enumerate}
    \item Convolutional layers: $\text{conv}(\mathbf{X}) = \{ \mathbf{A}^f \}_{f=1}^{F}$. Previously we used $\{ \mathbf{A}^f \}_{f=1}^{F}$ to denote convolutional feature maps from any layer; now we consider $\{ \mathbf{A}^f \}_{f=1}^{F}$ as the feature maps of specifically the last convolutional layer.
    
    \item One final fully connected layer: 
    \begin{align}
       \mathbf{s} = \mathbf{WA}+\mathbf{b},
    \end{align}
    where $\mathbf{s} \in \mathbb{R}^{M}$ are the raw scores for the $M$ classes, the weight matrix is $\mathbf{W} \in \mathbb{R}^{M \times F D_1 D_2}$, and $\mathbf{A} \in \mathbb{R}^{F D_1 D_2 \times 1}$ is the convolutional output $\{ \textbf{A}^f \}_{f=1}^{F}$ flattened. If we consider only a single class $m$, and extract out the relevant row of weights $\mathbf{w}_m$ and bias $b_m$, the expression for the $m^{th}$ class score is
    \begin{align}
        s_m = \mathbf{w}_m \mathbf{A} + b_m,
    \end{align}
    where $\mathbf{w}_m \in \mathbb{R}^{1 \times F D_1 D_2}$.
    
\end{enumerate}

Then $\frac{\partial s_m}{\partial \mathbf{A}} = \mathbf{w}_m$ and the HiResCAM explanation for class $m$ is

\begin{align}
\begin{split}
    \mathcal{\tilde{A}}_m^{\rm{HiResCAM}} &= \sum_{f=1}^{F} \frac{\partial s_m}{\partial \mathbf{A}} \odot \mathbf{A} \\ 
    &= \sum_{f=1}^{F} \mathbf{w}_m \odot \mathbf{A},
\end{split}
\end{align}

where here we consider $\mathbf{w}_m$ and $\mathbf{A}$ represented with the following dimensions: $\mathbf{w}_m \in \mathbb{R}^{F \times D_1 \times D_2}$ and $\mathbf{A} \in \mathbb{R}^{F \times D_1 \times D_2}$. In other words, we exploit the fact that $\mathbf{A}$ has spatial information and we can thus infer spatial information for the corresponding elements of $\mathbf{w}_m$.

HiResCAM highlights only relevant locations because $\mathbf{w}_m \odot \mathbf{A}$ in the HiResCAM calculation (where $\mathbf{w}_m \in \mathbb{R}^{F \times D_1 \times D_2}$ and $\mathbf{A} \in \mathbb{R}^{F \times D_1 \times D_2}$) is the intermediate computation in the input-specific part of the class score $\mathbf{w}_m \mathbf{A}$ (where $\mathbf{w}_m \in \mathbb{R}^{1 \times F D_1 D_2}$ and $\mathbf{A} \in \mathbb{R}^{F D_1 D_2 \times 1}$). 
\end{proof}

Figure \ref{fig:hirescamgradcamconcrete} provides a specific example demonstrating how HiResCAM explanations reflect the class score calculation, while Grad-CAM explanations do not.

\subsubsection{Grad-CAM can highlight irrelevant locations that are not guaranteed to increase the class score}

Grad-CAM does not directly visualize important locations, and Grad-CAM does not reflect the model's computations, even if Grad-CAM is applied at the last convolutional layer. Considering the 2D CNN described in the previous section, we begin calculating the Grad-CAM explanation by substituting in $\mathbf{w}_m$ for $\frac{\partial s_m}{\partial \mathbf{A}}$ in the importance weights equation \eqref{eqn:importanceweights}:
\begin{align} \label{eqn:gradcamavgwm}
    \alpha_m^f = \frac{1}{D_1 D_2} \sum_{d_1=1}^{D_1} \sum_{d_2=1}^{D_2} \mathbf{w}_m.
\end{align}
The final Grad-CAM explanation is then calculated by multiplying these importance weights $\alpha_m^f$ against the corresponding feature maps of $\mathbf{A}$. From equation \ref{eqn:gradcamavgwm} we see that all the model's fully connected layer weights corresponding to a given feature map are averaged together; averaging of the model's weights prevents the Grad-CAM explanation from reflecting the model's class score calculation.

\subsubsection{HiResCAM connection to regression}

A CNN ending in a single fully connected layer is a feature extractor (convolutions) followed by regression (the fully connected layer). The feature extractor produces a collection of numbers summarizing the input, and this collection of numbers is fed in to the regression to make the final prediction. Regression is fully interpretable. For a simple regression model $y = w_1 x_1 + w_2 x_2$, a \textit{global explanation} consists of inspecting the numeric values of the learned coefficients $w_1$ and $w_2$, while a \textit{local explanation} considers the values of $w_1 x_1$ and $w_2 x_2$ to understand the prediction for a particular input example. The global explanation is equivalent to the model gradient while the local explanation is equivalent to the gradient multiplied element-wise by the input \cite{ancona2019XAIBook}. HiResCAM is thus the local explanation that demonstrates for a particular example the most important locations for the prediction.

\subsection{CNNs ending in global average pooling then one fully-connected layer: the CAM architecture} \label{sec:camarch}

In the previous section, we considered CNNs that end in one fully-connected layer. In this section, we consider a narrower group of CNNs: those of the ``CAM architecture'' which end with global average pooling of feature maps followed by one fully-connected layer. By considering CAM architecture CNNs, we prove that HiResCAM is a generalization of CAM (Section \ref{sec:hiresgeneralizecam}), which then reveals that CAM, HiResCAM, and Grad-CAM applied at the last convolutional layer produce identical explanations for the CAM architecture (Section \ref{sec:allthesame}).

\subsubsection{Class Activation Mapping (CAM)}

Class Activation Mapping \cite{zhou2016CAM}, or CAM, is a CNN explanation method that requires the CAM architecture. CAM was developed with the observation that for this particular architecture, it is straightforward to visualize the locations in the input image that contribute to an increased score for a particular class.

Consider the following 2D CNN:

\begin{enumerate}
    \item Convolutional layers: $\text{conv}(\mathbf{X}) = \{ \textbf{A}^f \}_{f=1}^{F}$ where again here we consider $\{ \mathbf{A}^f \}_{f=1}^{F}$ as the feature maps of specifically the last convolutional layer.
    
    \item Global average pooling of each feature map to a scalar: \\ 
    \begin{align}
        a^f = \frac{1}{D_1 D_2} \sum_{d_1=1}^{D_1} \sum_{d_2=1}^{D_2} \textbf{A}_{d_1 d_2}^{f}. \label{eqn:gap}
    \end{align}
    
    \item Final fully connected layer: 
    \begin{align}
        s_m = w_m^1a^1 + w_m^2a^2 + \cdots + w_m^Fa^F \text{ for } m=1,...,M  \label{eqn:finalfc}
    \end{align}
    where $s_m$ is the score for the $m^{th}$ class and there are $M$ classes total. 
    
\end{enumerate}

The CAM explanation for class $m$ is then defined as:

\begin{align}
    & \mathcal{\tilde{A}}_m^{\rm{CAM}} = w_m^1\mathbf{A}^1 + w_m^2\mathbf{A}^2 + \cdots + w_m^F\mathbf{A}^F,
\end{align}

or equivalently:
\begin{align}
   &  \mathcal{\tilde{A}}_m^{\rm{CAM}} = \sum_{f=1}^{F} w_m^f\mathbf{A}^f. \label{eqn:cammaps}
\end{align}

\subsubsection{HiResCAM is a generalization of CAM} \label{sec:hiresgeneralizecam}

In this section we demonstrate that HiResCAM (equation \ref{eqn:hirescam}) is a generalization of CAM. 

\begin{proof}

As before, to calculate the HiResCAM explanation for class $m$, we first calculate $\frac{\partial s_m}{\partial \mathbf{A}^f}$. To get a useful expression for $\frac{\partial s_m}{\partial \mathbf{A}^f}$ in a CAM architecture, we plug in the right-hand side of the global average pooling equation \ref{eqn:gap} into the final fully connected layer equation \ref{eqn:finalfc}:

\begin{align}
\begin{split}
     s_m &= w_m^1(\frac{1}{D_1 D_2}\sum_{d_1=1}^{D_1} \sum_{d_2=1}^{D_2} \mathbf{A}_{d_1 d_2}^1) \\
         &+ w_m^2(\frac{1}{D_1 D_2}\sum_{d_1=1}^{D_1} \sum_{d_2=1}^{D_2} \mathbf{A}_{d_1 d_2}^2)  \\
         &+ \cdots + w_m^F(\frac{1}{D_1 D_2}\sum_{d_1=1}^{D_1} \sum_{d_2=1}^{D_2} \mathbf{A}_{d_1 d_2}^F).
\end{split}
\end{align}

From the above expression we can see that the gradient $\frac{\partial s_m}{\partial \mathbf{A}^f}$ is\footnote{For another perspective on equation \ref{eqn:camgradient}, consider that the gradient of the score for class $m$ with respect to the pooling outputs $a^1, a^2,...,a^F$ are the respective $m$-specific weights $w_m^1, w_m^2,...,w_m^F$. Each $w_m^f$ value is then propagated back through the global average pooling step by considering how many elements were pooled together, and distributing $w_m^f$ equally across all those elements - that is, dividing $w_m^f$ by $D_1 D_2$, which is the total number of elements in each feature map.}

\begin{align} \label{eqn:camgradient}
    & \frac{\partial s_m}{\partial \mathbf{A}^f} = \frac{1}{D_1 D_2} w_m^f,
\end{align}

which means that the HiResCAM explanation for a model with a CAM architecture is:

\begin{align}
     \mathcal{\tilde{A}}_m^{\rm{HiResCAM}} &= \frac{1}{D_1 D_2} \sum_{f=1}^{F} w_m^f \mathbf{A}^f.
\end{align}

This is identical to the CAM explanation (equation \ref{eqn:cammaps}) except for a constant factor of $\frac{1}{D_1 D_2}$ which disappears in the subsequent normalization step.

Thus, HiResCAM is a generalization of CAM because both methods yield identical explanations for any CAM architecture, but HiResCAM is applicable to a broader class of architectures as shown previously in Section \ref{sec:cnn-end-one-fc}.
\end{proof}

\subsubsection{CAM, HiResCAM, and Grad-CAM produce identical explanations for CNNs of the CAM architecture} \label{sec:allthesame}

Previous work \cite{selvaraju2017grad} has demonstrated that Grad-CAM is also a generalization of CAM. Thus, HiResCAM and Grad-CAM are alternative generalizations of the CAM method, and for the CAM architecture the following methods produce identical explanations: CAM, HiResCAM applied at the last convolutional layer, and Grad-CAM applied at the last convolutional layer. Examples of CNNs following the CAM architecture include ResNets \cite{he2016deep}, GoogLeNet \cite{szegedy2015going}, and DenseNets \cite{huang2017densely}.

\subsection{All other CNNs}

Section \ref{sec:cnn-end-one-fc} considered CNNs ending in one fully connected layer, while Section \ref{sec:camarch} considered the narrower class of CAM architecture CNNs. For all other CNNs, including those ending in multiple fully connected layers, CAM explanations cannot be calculated, and while HiResCAM and Grad-CAM explanations can be calculated, they are not provably guaranteed to highlight only relevant locations. To the best of our knowledge, no gradient-based neural network explanation method yet exists which can produce consistently class-specific explanations guaranteed to highlight only the locations the model is using, for any arbitrary CNN architecture or layer (see Related Work). For sensitive applications where trustworthy class-specific explanations are required, we recommend using CAM with a CAM architecture CNN, or HiResCAM with any CNN ending in only one fully connected layer.

\subsection{HiResCAM connection to Gradient $*$ Input} 

The preceding sections demonstrated HiResCAM's relationship to CAM, Grad-CAM, and regression. HiResCAM is also related to Gradient $*$ Input. If HiResCAM were to be applied at the level of the input image, it would be equivalent to Gradient $*$ Input. While the ``level of application'' is a simple distinction, it has several important implications. First, HiResCAM explanations are clean and class-specific, whereas Gradient $*$ Input and other pixel-space explanations produce visualizations that are too noisy to be class-specific \cite{selvaraju2017grad}. Second, HiResCAM is not susceptible to the same issue that caused Gradient $*$ Input to fail sanity checks \cite{adebayo2018sanity}: namely, HiResCAM does not involve element-wise multiplication with the raw input image. Third, when HiResCAM is used as recommended, the HiResCAM explanations can be seamlessly integrated into model training in a computationally efficient manner, as the gradients necessary to compute the HiResCAM explanation correspond to particular model weights and can thus be accessed during the forward pass without any extra backward passes required.
%
%

\section{Model explanation is not weakly-supervised segmentation} \label{sec:notweaksupseg}

There are close ties between model explanation and weakly-supervised segmentation (WSS), but these tasks have different goals. This section provides context for the Experiments section by assessing the relationship between model explanation and WSS.

\subsection{The goal of model explanation}

The goal of model explanation is to demonstrate what locations in an image a model used to make a particular prediction - \textit{even if that means highlighting areas outside of the relevant object}. For example, if a model has used tracks to identify a train, the explanation should highlight the tracks. If the model has used water to identify a boat, the explanation should highlight the water. Any performance metric to evaluate explanation correctness (``explanation quality'') thus must be calculated against a ground truth of ``locations the model used for each prediction'' which in turn can only be uncovered through mathematical properties of a model and an explanation method - for example, by proving that HiResCAM exactly explains locations used by CNNs ending in one fully connected layer. Locations the model used may not have any relation to object segmentation maps, nor can they be created manually by a human, for if humans were able to understand models well enough to circumscribe regions used for each prediction then there would be no need for explanation methods in the first place.

\subsection{The goal of weakly-supervised segmentation}
The goal of weakly-supervised segmentation is to identify all pixels that are part of an object using only a classification model trained on whole image labels. Because classifiers tend to focus on small discriminative parts of objects \cite{lee2019ficklenet, li2018GAIN, chang2020SCCAM, kolesnikov2016SEC}, a key goal of weakly-supervised segmentation methods is to expand the attention of the classifier beyond small discriminative areas so that the attention covers more of the object, such as by using conditional random fields \cite{chang2020SCCAM}. 

Somewhat confusingly, many methods for weakly-supervised segmentation are based on the model explanation methods CAM or Grad-CAM and thus may be named similarly, including Mixup-CAM \cite{chang2020mixupcam}, Sub-Category CAM \cite{chang2020SCCAM}, Puzzle-CAM \cite{jo2021puzzle}, and FickleNet \cite{lee2019ficklenet}. However, it is critical to keep in mind that even though these WSS approaches are leveraging model explanations, they have a fundamentally different goal. 

\subsection{IoU must not be used to evaluate explanation correctness}

One common performance metric for WSS is intersection-over-union (IoU), which is highest when the predicted segmentation for a class fully overlaps the ground truth segmentation for that class without spreading to other regions. While IoU is a reasonable metric for judging WSS performance, IoU should never be used to evaluate explanation correctness. Unfortunately, some prior work attempts to experimentally evaluate explanation correctness using IoU calculated against ground truth object segmentations, or using the closely related setup of asking humans to subjectively judge how well explanations correspond to an object (which in effect corresponds to humans estimating an IoU-like quantity). The unspoken assumption in these experiments is that the classification model is always using the relevant object to predict the class and so a ``good'' explanation method will achieve a high IoU. However, as mentioned previously, models are \textit{not} guaranteed to always use the relevant object to predict the class, and indeed, the possibility for undesirable model behavior is a primary motivator behind development of explanation methods. Any time a model behaves unexpectedly or exploits spurious correlations, the IoU of a truthful model explanation will be low, but it would be false to then conclude that the low IoU means the explanation was of poor quality. The only way to know if an explanation method is faithful to a particular type of model is to prove it.

\subsection{IoU of a faithful explanation method provides insight into a model}

Although IoU cannot be used to evaluate the quality of an explanation method, IoU calculated based on an explanation method with guaranteed properties (\textit{e.g.} CAM or HiResCAM) can be used to evaluate a particular \textit{model}. In these cases, high IoU indicates that the model tends to make predictions using areas within the objects of interest, as desired, while low IoU indicates that the model tends to make predictions using areas outside the object of interest and thus is exploiting background or correlated objects.

\subsection{HiResCAM is an explanation method}

The goal of HiResCAM is not to expand the size of the attention maps or to yield high performance on weakly-supervised segmentation. HiResCAM is an explanation method that faithfully represents the locations the model has used to make a prediction, even if these locations are outside the object of interest.

\section{Experiments}

\subsection{Datasets}

To compare HiResCAM and Grad-CAM, in Sections \ref{sec:l2-distance-corr} through \ref{sec:amt-human} we conduct experiments on PASCAL VOC 2012 \cite{everingham2010pascal}, a data set of 2D RGB natural images with ground truth segmentation maps of 20 classes. Following prior work \cite{wang2020SEAM}, we combine PASCAL VOC 2012 with SBD \cite{hariharan2011semantic} to create an augmented training set of 7,087 images. Results are reported on the PASCAL VOC 2012 validation set which contains 1,449 images annotated with 2,148 segmentation maps. 

In Section \ref{sec:ct-experiments} we provide a qualitative comparison of HiResCAM and Grad-CAM on the medical imaging dataset RAD-ChestCT, which includes 36,316 chest computed tomography scans annotated with 83 different abnormalities \cite{draelos2021machine}.

\subsection{Models}

For the experiments on natural images we report results for a ResNet-34 variant \cite{he2016deep} and a DenseNet-121 variant \cite{huang2017densely} which both end in one fully connected layer and avoid global average pooling, so that HiResCAM and Grad-CAM do not collapse to CAM. These models are defined in the Appendix. Both models use convolutional layers pretrained on ImageNet \cite{deng2009imagenet}. All layers are refined using only whole-image labels from PASCAL VOC 2012. PASCAL VOC 2012 experiments were conducted on an NVIDIA Titan RTX GPU with 24 GB of memory. Results on the RAD-ChestCT dataset are from an AxialNet model \cite{draelos2021explainable} and were obtained using an NVIDIA Tesla V100 GPU with 32 GB of memory.

\subsection{Code}

All code to replicate all experiments will be made publicly available on GitHub following publication.

\subsection{Explanation correctness: HiResCAM reflects model computations while Grad-CAM does not} \label{sec:l2-distance-corr}

We first quantify the explanation correctness of HiResCAM and Grad-CAM on PASCAL VOC 2012 by calculating the L2 distance between the explanation and the ground truth of the locations the model used. The models are CNNs ending in a single fully connected layer, meaning that the HiResCAM explanation is provably identical to the locations the model used; therefore, the calculated L2 distance between the HiResCAM explanation and the ground truth of the locations used is always 0. However, for Grad-CAM, the explanation is \textit{not} equivalent to the locations used, and so the distance is always nonzero, demonstrating that Grad-CAM provides misleading explanations that do not illustrate the model's behavior (Table \ref{tab:l2togrtruth}).

\begin{table}
\centering
\caption{Explanation correctness evaluated by mean L2 distance between the explanation and the ground truth of the locations each model used to make predictions on the PASCAL VOC 2012 validation set. An L2 distance of 0 indicates that the explanation is correct and exactly reflects the locations the model used. A nonzero L2 distance indicates that the explanation is misleading and does not accurately reflect the model's computations.}
\label{tab:l2togrtruth}
\begin{tabular}{@{}lcccc@{}}
& \multicolumn{4}{c}{\textbf{L2 Distance to Locations Model Used}} \\
\toprule
      & \multicolumn{2}{c}{\textbf{DenseNet-121v}} & \multicolumn{2}{c}{\textbf{ResNet-34v}} \\ \cmidrule{2-3} \cmidrule{4-5}
\textbf{Label} & \textbf{Grad-CAM} & \textbf{HiResCAM} & \textbf{Grad-CAM} & \textbf{HiResCAM} \\ \cmidrule{1-5}
airplane & 2.80 & 0 & 3.03 & 0 \\
bicycle & 1.02 & 0 & 0.96 & 0 \\
bird & 2.76 & 0 & 2.69 & 0 \\
boat & 0.99 & 0 & 0.96 & 0 \\
bottle & 0.85 & 0 & 0.81 & 0 \\
bus & 0.91 & 0 & 0.90 & 0 \\
car & 0.72 & 0 & 1.05 & 0 \\
cat & 2.24 & 0 & 2.34 & 0 \\
chair & 0.72 & 0 & 0.74 & 0 \\
cow & 0.88 & 0 & 1.12 & 0 \\
dining table & 0.77 & 0 & 0.85 & 0 \\
dog & 1.47 & 0 & 2.57 & 0 \\
horse & 1.06 & 0 & 1.29 & 0 \\
motorbike & 1.11 & 0 & 1.28 & 0 \\
person & 0.95 & 0 & 1.08 & 0 \\
potted plant & 0.70 & 0 & 0.64 & 0 \\
sheep & 1.20 & 0 & 1.06 & 0 \\
sofa & 0.71 & 0 & 0.63 & 0 \\
train & 1.12 & 0 & 1.56 & 0 \\
tv monitor & 0.75 & 0 & 0.71 & 0 \\ \bottomrule
\end{tabular}
\end{table}

\subsection{Explanation utility for weakly supervised segmentation}

As detailed in Section \ref{sec:notweaksupseg}, weakly supervised segmentation (WSS) cannot be used to evaluate an explanation's correctness. However, as CAM explanations are frequently used as a starting point for WSS methods, this section provides WSS performance of raw HiResCAM and Grad-CAM explanations.

To calculate IoU, an explanation with continuous values in the range $[0,1]$ must be binarized into a proposed segmentation map with values in the set $\{0,1\}$. Choosing a poor binarization threshold can lead to a low IoU that is not reflective of how well the explanation overlaps the object. To ensure a fair comparison, we therefore selected the best binarization threshold for each explanation method, object class, and model separately, considering thresholds between 0.02 and 0.98 in increments of 0.02. We did not apply any post-processing techniques such as region growing to avoid confounding the results.

The results are summarized in Table \ref{tab:resnet34-ioubylabel}. Interestingly, Grad-CAM outperforms HiResCAM at WSS. We hypothesize that this may be because Grad-CAM tends to ``expand'' explanations beyond the regions the model used, an idea that we investigate further in the next section.

Because the HiResCAM explanation reflects the model's computations, we can additionally use the HiResCAM IoU to understand the extent to which these particular models predict each class based on the object itself versus background or other correlated objects. As is apparent from the relatively low IoUs, the models appear to be making some use of background and correlated objects in their predictions.

\begin{table}
\centering
\caption{Explanation utility for weakly supervised segmentation: mean IoU calculated for class-specific explanations versus the corresponding class-specific ground truth segmentation maps provided in the PASCAL VOC 2012 validation set. Note that the primary purpose of HiResCAM is accurate explanation (Table \ref{tab:l2togrtruth}), not WSS. The ``overall'' row at the bottom was calculated by averaging IoU across all image-class pairs, so classes appearing more frequently are weighted proportionally more.}
\label{tab:resnet34-ioubylabel}
\begin{tabular}{@{}lcccc@{}}
& \multicolumn{4}{c}{\textbf{Mean IoU}} \\
\toprule
      & \multicolumn{2}{c}{\textbf{DenseNet-121v}} & \multicolumn{2}{c}{\textbf{ResNet-34v}} \\ \cmidrule{2-3} \cmidrule{4-5}
\textbf{Label} & \textbf{Grad-CAM} & \textbf{HiResCAM} & \textbf{Grad-CAM} & \textbf{HiResCAM} \\ \cmidrule{1-5}
airplane & 30.6 & 35.5 & 36.0 & 36.0 \\
bicycle & 13.9 & 14.3 & 15.3 & 15.7 \\
bird & 25.4 & 31.4 & 29.4 & 30.9 \\
boat & 20.8 & 19.8 & 24.1 & 23.4 \\
bottle & 19.4 & 19.0 & 20.2 & 19.0 \\
bus & 40.5 & 41.8 & 48.3 & 45.6 \\
car & 30.1 & 27.8 & 34.1 & 27.8 \\
cat & 39.6 & 39.5 & 42.5 & 41.0 \\
chair & 15.4 & 12.4 & 15.4 & 12.6 \\
cow & 37.8 & 34.9 & 36.9 & 37.5 \\
dining table & 31.5 & 28.4 & 31.5 & 28.4 \\
dog & 36.3 & 38.4 & 38.0 & 38.5 \\
horse & 27.7 & 30.6 & 35.3 & 35.3 \\
motorbike & 36.8 & 37.7 & 42.8 & 41.3 \\
person & 29.9 & 28.4 & 29.4 & 23.5 \\
potted plant & 19.2 & 14.8 & 19.5 & 15.0 \\
sheep & 36.2 & 37.1 & 35.2 & 34.6 \\
sofa & 32.0 & 26.3 & 36.3 & 28.3 \\
train & 32.6 & 34.1 & 41.5 & 43.0 \\
tv monitor & 35.2 & 31.1 & 38.4 & 33.4 \\ \cmidrule{1-5}
overall & \textbf{29.4} & 28.9 & \textbf{31.8} & 29.1 \\ \bottomrule
\end{tabular}
\end{table}

\subsection{Humans perceive Grad-CAM and HiResCAM explanations differently} \label{sec:amt-human}

\begin{figure}
    \centering
    \includegraphics[scale=0.6]{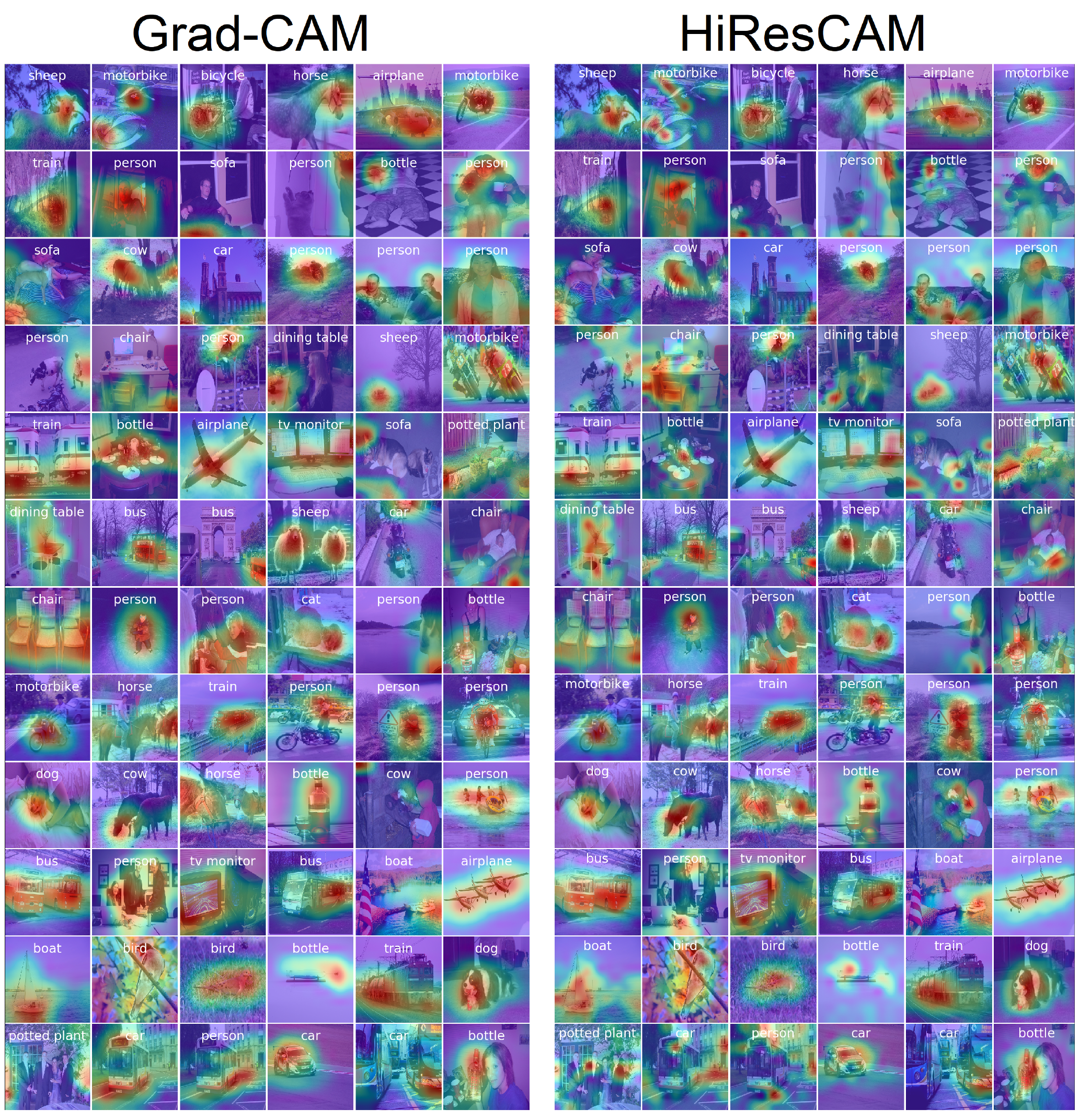}
    \caption{HiResCAM and Grad-CAM explanations for 72 randomly-selected PASCAL VOC 2012 validation set image-class pairs for ResNet-34v. Each half of the figure includes the same images and classes. The two halves differ only in the calculation of the explanation. Best viewed in color.}
    \label{fig:additionalPASCAL}
\end{figure}

\begin{figure}
    \centering
    \includegraphics[scale=0.25]{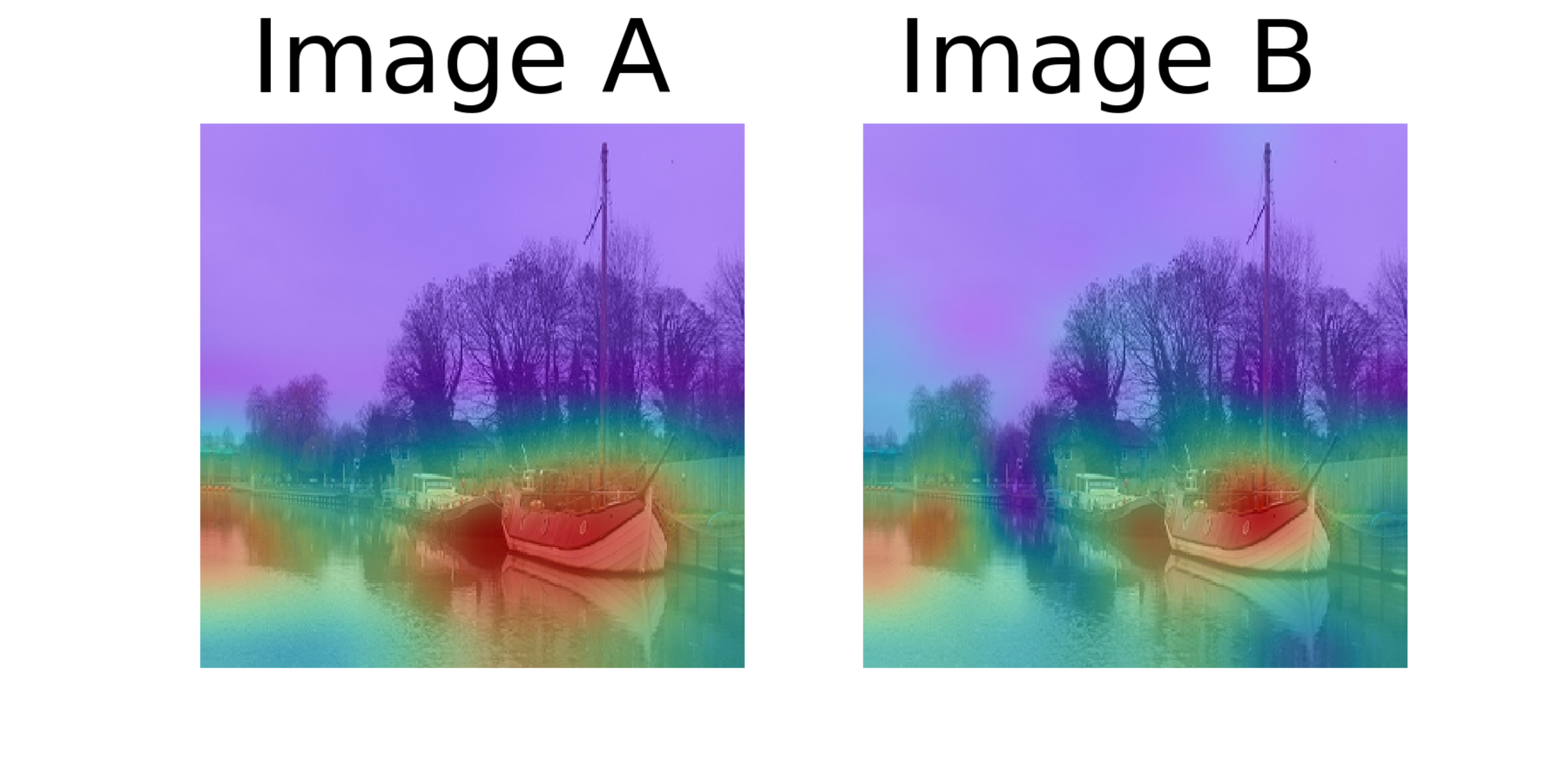}
    \caption{Example of a visualization used in the AMT human evaluation task comparing Grad-CAM and HiResCAM explanations produced using the same model on the same input image and class. HiResCAM appeared as Image A in 50\% of the visualizations and as Image B in the remaining 50\%. Best viewed in color.}
    \label{fig:amt-setup-boat-pic}
\end{figure}

Figure \ref{fig:additionalPASCAL} displays HiResCAM and Grad-CAM explanations for 72 randomly-selected image-class pairs. Inspection of this figure suggests that Grad-CAM may be expanding attention maps. In order to quantify how human perception of HiResCAM and Grad-CAM explanations differ, we ran Amazon Mechanical Turk (AMT) experiments. We selected 250 image-class pairs randomly from the PASCAL VOC 2012 validation set and plotted the ResNet-34v Grad-CAM explanation next to the HiResCAM explanation such that HiResCAM sometimes randomly appeared on the left as ``Image A'' and other times randomly appeared on the right as ``Image B''  (Figure \ref{fig:amt-setup-boat-pic}). Workers were never informed which explanation method appeared on the right versus the left. Workers were asked three multiple-choice questions to compare the two explanations:

\begin{itemize}
    \item A size question, with mutually exclusive options ``Image A highlight is bigger;'' ''Image B highlight is bigger;'' ``Image A and B highlights are the same size.''
    \item A shape question, with mutually exclusive options ``Image A highlight is smoother/rounder;'' ``Image B highlight is smoother/rounder;'' ``Image A and B highlights have similar shapes''
    \item A focus question, with mutually exclusive options ``Image A highlight is more focused on the \textit{label};'' ``Image B highlight is more focused on the \textit{label};'' ``Image A and B highlights have similar focus on the \textit{label}'' where \textit{label} was replaced with the particular class being explained, e.g. ``person'' or ``tv monitor.'' If neither Image A nor Image B was focused on the object, the workers were instructed to choose the ``similar focus'' option. 
\end{itemize}

For each of the three questions for the 250 paired explanations, five unique workers provided an answer, for a total of 3,750 human evaluations. The total number of workers who participated was 42. Worker quality was assessed by calculating how often a worker's answer agreed with the mode(s) of the other 4 answers provided per question.

\begin{figure}
    \centering
    \includegraphics[scale=0.7]{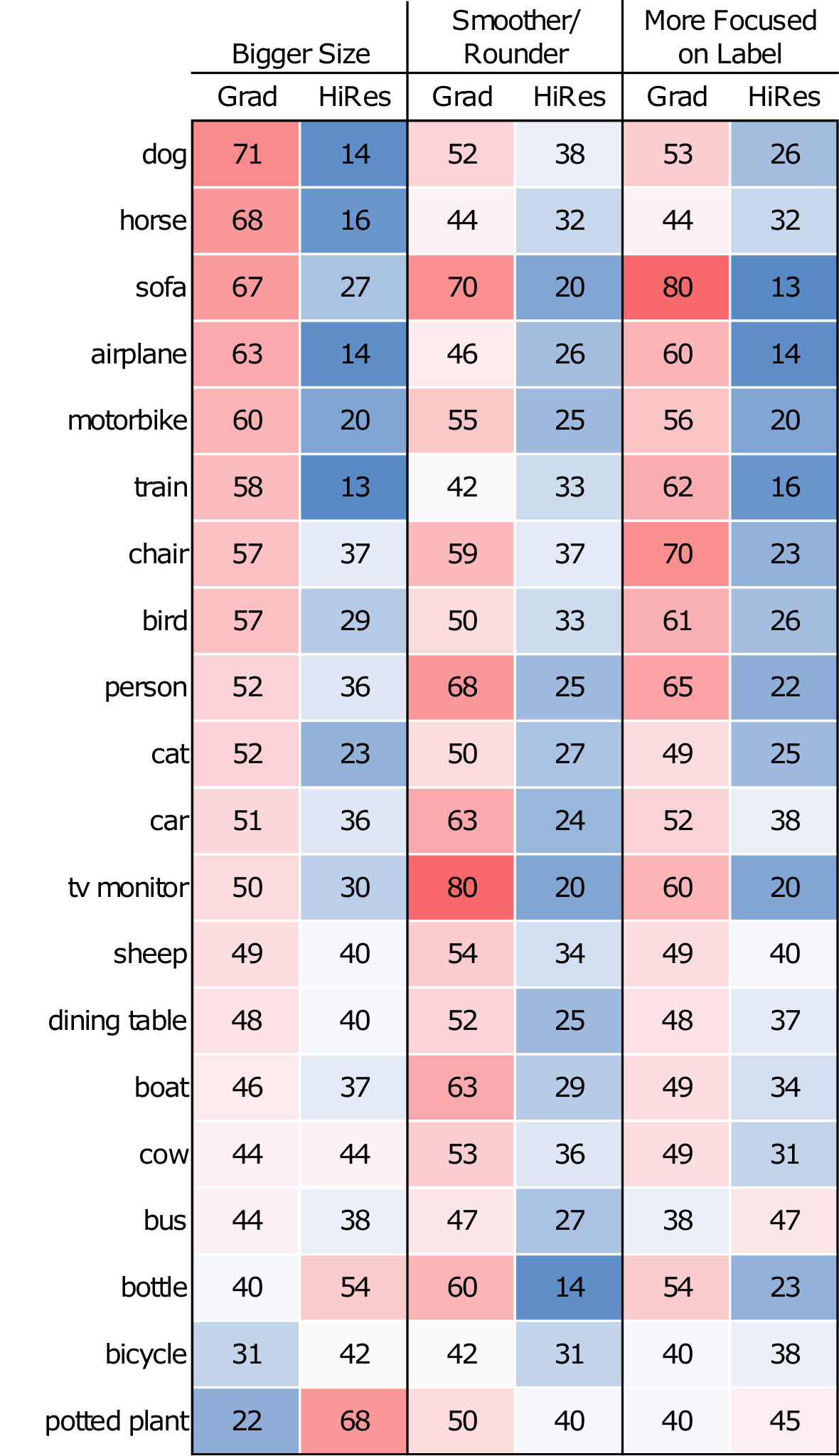}
    \caption{Amazon Mechanical Turk human evaluation results comparing ResNet-34v Grad-CAM and HiResCAM explanations on size, shape, and focus. Considering the ``Bigger Size'' comparison as an example, the ``Grad'' column indicates the percent of the time that workers judged the Grad-CAM explanation to be bigger in size than HiResCAM, while the ``HiRes'' column indicates the percent of the time workers judged HiResCAM to be bigger. The percents do not add up to 100 across the Grad and HiRes columns because for each characteristic workers were also allowed to indicate that Grad-CAM and HiResCAM were equivalent. For most classes, workers perceived Grad-CAM explanations as bigger, smoother/rounder, and more focused on the relevant object. Best viewed in color.}
    \label{fig:amt-sizeshapefocus}
\end{figure}

AMT results are shown in Figure \ref{fig:amt-sizeshapefocus}. Although we demonstrated previously that Grad-CAM does not reflect the model's computations and therefore should not be used for model explanation, the results of the AMT task intriguingly suggest that Grad-CAM may have extra utility for WSS, as the Grad-CAM explanations are typically bigger, smoother/rounder, and more focused on the object than the HiResCAM explanations.

\begin{figure}
    \centering
    \includegraphics[scale=0.11]{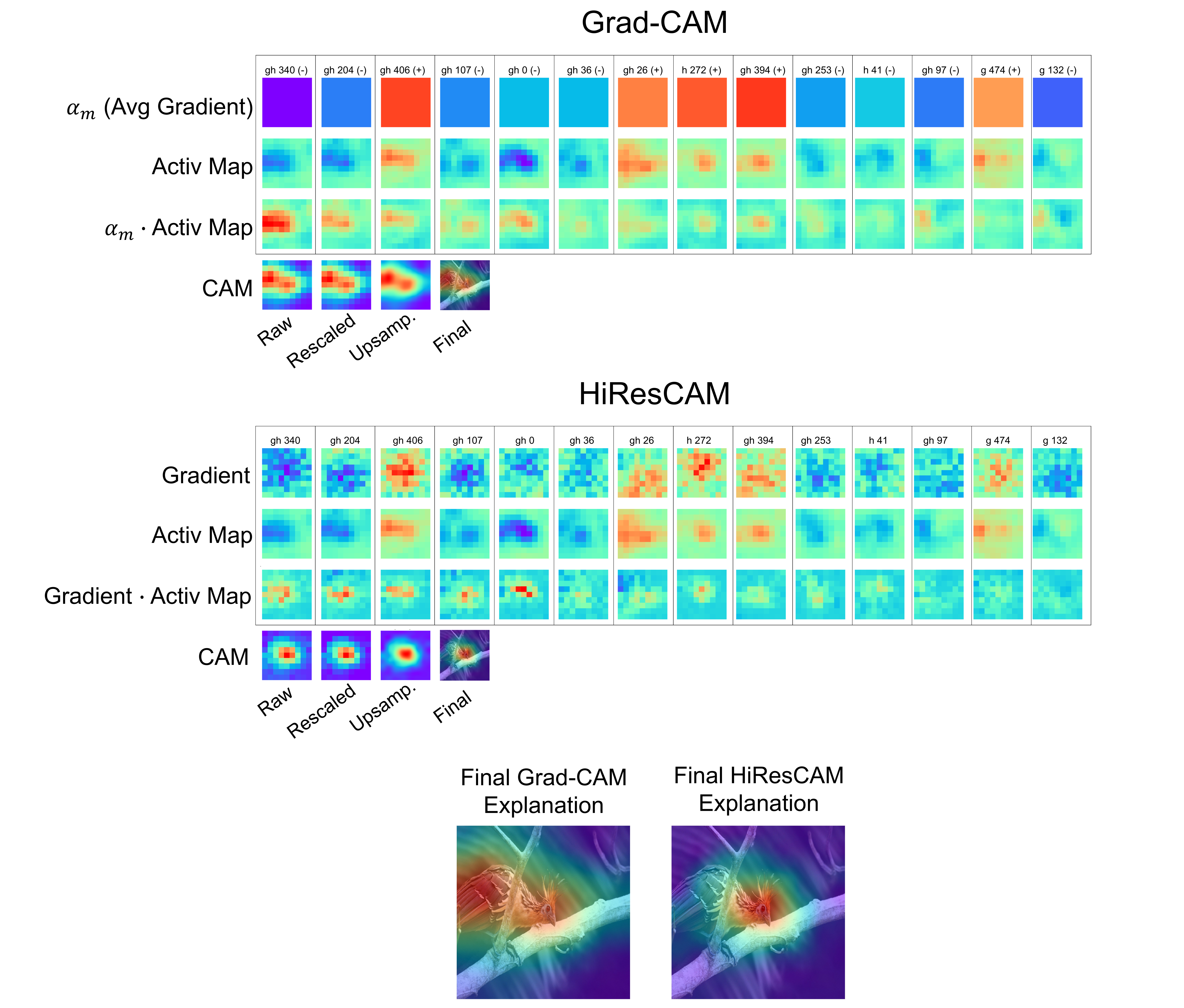}
    \caption{Step-by-step example of how Grad-CAM and HiResCAM explanations are calculated, using a real PASCAL VOC 2012 validation set image with gradients and activation maps from ResNet-34v. The Grad-CAM explanation has expanded the attention beyond the locations the model used by blurring out the gradient information and thus placing more emphasis on certain features that in this case happen to cover more of the object. Figure arrangement: Each boxed column corresponds to one feature dimension. As this model ends in 512 features, for human comprehensibility only the union of the top 12 features that contribute most to each final explanation are shown. Because there was overlap between the top 12 features for HiResCAM and the top 12 features for Grad-CAM for this image, a total of 14 unique features are included. Further explanation of this figure's construction is provided in the Appendix. Best viewed in color, with zoom.}
    \label{fig:stepbystepbird}
\end{figure}

To better understand the manner in which Grad-CAM explanations expand beyond the locations the model used, we generated step-by-step examples showing the intermediate calculations for HiResCAM and Grad-CAM explanations. One such example is shown in Figure \ref{fig:stepbystepbird}. This example suggests that Grad-CAM may expand attention maps by over-emphasizing top feature maps. 

\subsection{In medical images, Grad-CAM can create the incorrect impression that the model has focused on the wrong organ} \label{sec:ct-experiments}

\begin{figure}
    \centering
    \includegraphics[scale=0.10]{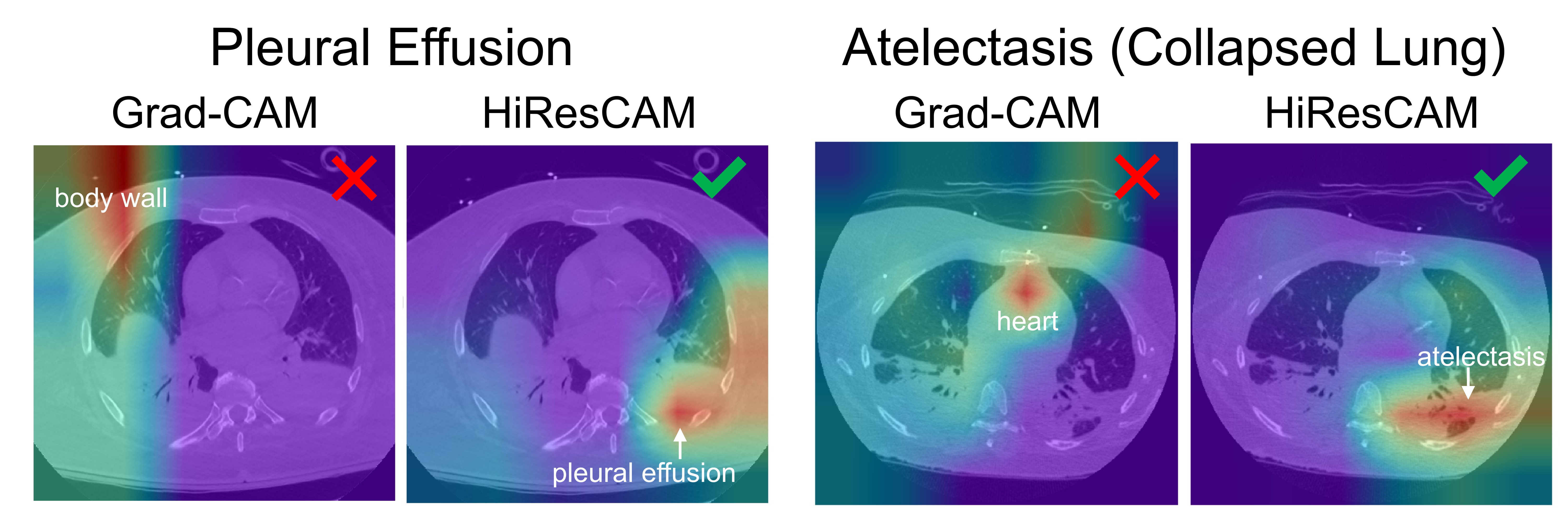}
    \caption{Examples of Grad-CAM creating the incorrect impression that an AxialNet model focused on the wrong anatomical structure. The HiResCAM and Grad-CAM explanations were generated using exactly the same model on the same input CT volume. The only difference is the explanation method. Both of the abnormalities shown here are lung findings, and HiResCAM indicates that the model used the lung fields to predict these lung findings. However, Grad-CAM creates the impression that the model predicted these lung abnormalities based on the body wall and heart, which are irrelevant. Best viewed in color; text annotations added for clarity.}
    \label{fig:NewFigure1CTExample}
\end{figure}

Reliable model explanation is particularly important for sensitive applications, such as those in criminal justice or healthcare. Concurrent work \cite{draelos2021explainable} assessed HiResCAM and Grad-CAM for explainable multiple abnormality prediction in volumetric medical images. We leveraged that final AxialNet model to generate additional HiResCAM and Grad-CAM visualizations that reveal how Grad-CAM sometimes creates the false impression that the model focused on the wrong anatomical structure when predicting an abnormality (Figure \ref{fig:NewFigure1CTExample}). For a more detailed comparison of HiResCAM and Grad-CAM in medical imaging, see \cite{draelos2021explainable}.

\section{Conclusion}

In this work, we propose the model explanation method HiResCAM, a new generalization of CAM that is provably guaranteed to highlight locations used by any CNN ending in one fully connected layer. We demonstrate that the related method Grad-CAM can create misleading explanations, but because it also tends to expand attention maps, Grad-CAM yields superior WSS performance. Overall, Grad-CAM's attention expansion properties may be useful for downstream segmentation tasks, while for model understanding HiResCAM provides faithful, class-specific explanations.

\section*{Acknowledgements}

The authors would like to thank Paidamoyo Chapfuwa, Ph.D., for thoughtful comments on a previous version of the manuscript, and Geoffrey D. Rubin, M.D., FACR, for helpful remarks on the medical imaging application.

\section*{Funding Sources}

This work was supported in part by the National Institutes of Health (NIH) Duke Medical Scientist Training Program Training Grant (GM-007171).


\bibliography{Sources}

\section{Appendix}

\subsection{ResNet-34v and DenseNet-121v Models}

In the main paper we report results on PASCAL VOC 2012 for two models, ResNet-34v and DenseNet-121v. The ResNet-34 variant (ResNet-34v) begins with all the standard ResNet-34 convolutional layers pretrained on ImageNet. The DenseNet-121 variant (DenseNet-121v) begins with all the standard DenseNet-121 convolutional layers pretrained on ImageNet. Both models end with a randomly-initialized custom convolutional layer (512 output feature maps, 2$\times$2 kernel, 1$\times$1 stride, no padding) followed by a randomly-initialized fully connected layer that produces the final predictions. All explanations were calculated at the last convolutional layer. All layers were refined during training on the PASCAL VOC 2012 classification task. The models do not include the global average pooling of the CAM architecture because otherwise HiResCAM and Grad-CAM both reduce to CAM and no meaningful comparison between the methods can be made.

\subsection{Details of the step-by-step figure}

Figure \ref{fig:stepbystepbird} shows a step-by-step example of how Grad-CAM and HiResCAM explanations were calculated for a PASCAL VOC 2012 validation set image and a ResNet-34v model. This section provides additional details explaining the figure. 

The figure illustrates only the top features for Grad-CAM and HiResCAM. For Grad-CAM, top features are defined as those for which the average value of ($\alpha_m \times$ activation map) is highest. For HiResCAM, top features are those for which the average value of (gradient $\times$ activation map) is highest. 

In the figure, a feature column is titled with a ``g'' if it was one of Grad-CAM's top features, an ``h'' if it was one of HiResCAM's top features, and a ``gh'' if it was a top feature for both explanation methods. For example, the leftmost feature in Figure \ref{fig:stepbystepbird} is ``gh 340'' meaning this is feature 340/512 and was a top contributor to both Grad-CAM and HiResCAM for this image and class. For Grad-CAM, the``(-)'' part of ``gh 340 (-)'' indicates that the $\alpha_m$ value was negative for this feature. 

The CAM row at the bottom shows the raw CAM first (the sum over the feature dimension of $\alpha_m \times$ activation map or gradient $\times$ activation map; all 512 features, even those not shown in the figure, are included in the CAM), then the CAM rescaled to [0,1] which has the same relative colors when visualized, then the CAM upsampled to the dimensions of the input image, and last the final explanation which is the upsampled CAM overlaid on the input image itself.

\subsection{The bias term of the final fully connected layer has no effect on CAM visualizations} \label{sec:biaseffect}

This section includes a miscellaneous observation about CAM explanations that was noted during the course of writing this paper. The authors of the CAM paper state, ``we ignore the bias term: we explicitly set the input bias of the softmax to 0 as it has little to no impact on the classification performance.'' We found that there is no need to explicitly ignore the bias term because it drops out of the visualization on its own. Section \ref{sec:hiresgeneralizecam} demonstrated that HiResCAM is a generalization of CAM. Calculating the CAM map from the HiResCAM perspective illustrates why the bias term is irrelevant to the explanation in a CAM architecture. Consider the score $s_m$ written in terms of the feature maps, for a final fully connected layer that does have a nonzero bias term:

\begin{align}
\begin{split}
    s_m &= w_m^1(\frac{1}{D_1 D_2}\sum_{d_1=1}^{D_1} \sum_{d_2=1}^{D_2} \mathbf{A}_{d_1 d_2}^1) \\
         &+ w_m^2(\frac{1}{D_1 D_2}\sum_{d_1=1}^{D_1} \sum_{d_2=1}^{D_2} \mathbf{A}_{d_1 d_2}^2)  \\
         &+ \cdots + w_m^F(\frac{1}{D_1 D_2}\sum_{d_1=1}^{D_1} \sum_{d_2=1}^{D_2} \mathbf{A}_{d_1 d_2}^F) \textcolor{red}{+ b_m}.
\end{split}
\end{align}

When calculating the gradient $\frac{\partial s_m}{\partial \mathbf{A}}$, the bias term disappears, because it has nothing to do with $\mathbf{A}$. Thus, the formula for CAM (or HiResCAM) remains the same, with or without a bias term.

There is some intuition about why the bias term of the final fully connected layer should not affect the visualizations: the point of CAM is to show an explanation for a particular class and a particular input image. However, the bias term will be the same for every image in an arbitrary collection of images, because the bias term is part of the model and is not input-dependent. Therefore, the bias term must not contribute anything input-specific to the explanation.


\end{document}